\documentclass[10pt,letterpaper,nofonttune]{IEEEtran}

\newcommand{\states}{\mathcal{S}}
\newcommand{\actions}{\mathcal{A}}

\newcommand{\buffer}{\mathcal{B}}

\usepackage[numbers,sort&compress]{natbib}
\usepackage{url}  
\usepackage{amsthm}
\usepackage{amssymb}
\usepackage{amsfonts}
\usepackage{amsmath}
\usepackage{eso-pic}
\usepackage{graphicx}
\usepackage{tikz}
\usepackage{pgf}
\usepackage{lscape}
\usepackage{glossaries}
\usepackage{fancyvrb}
\usepackage{verbatim}
\usepackage{xfrac}
\usepackage{algpseudocode,algorithm,algorithmicx}
\usepackage{multirow}
\usepackage{proof}
\usepackage{epstopdf}
\usepackage{subfigure}
\usetikzlibrary{arrows,automata}
\usetikzlibrary{shapes}
\newcommand{\reffig}[1]{Fig.~\ref{#1}}   

\theoremstyle{plain}

\newtheorem*{corollary*}{Corollary}

\theoremstyle{definition}

\newtheorem*{definition*}{Definition}

\theoremstyle{remark}

\newtheorem*{remark*}{Remark}

\title{\textit{Can You Fix My Neural Network?} \\Real-Time Adaptive Waveform Synthesis for Resilient Wireless Signal Classification}

\author{\IEEEauthorblockN{
Salvatore D'Oro\IEEEauthorrefmark{1},
Francesco Restuccia\IEEEauthorrefmark{1}\IEEEauthorrefmark{2},
and Tommaso Melodia\IEEEauthorrefmark{1}}\\
\IEEEauthorblockA{\IEEEauthorrefmark{1}Institute for the Wireless Internet of Things, Northeastern University, United States \\
\IEEEauthorrefmark{2}Roux Institute, Northeastern University, United States
\\ Email: \{s.doro, f.restuccia, t.melodia\}@northeastern.edu\vspace{-0.8cm}}\\
\thanks{This  work  was supported in part by NSF under grants CNS-1923789 and CCF-1937500 and in part by DARPA under grant HR00112090055. Visit ece.northeastern.edu/wineslab/index.php for more.}
\thanks{This paper has been accepted for publication in IEEE INFOCOM 2019. This is a preprint version of the accepted paper. Copyright (c) 2013 IEEE. Personal use of this material is permitted. However, permission to use this material for any other purposes must be obtained from the IEEE by sending a request to pubs-permissions@ieee.org.}
}

\newacronym{6g}{6G}{sixth generation}
\newacronym{3gpp}{3GPP}{3rd Generation Partnership Project}
\newacronym{adc}{ADC}{Analog to Digital Converter}
\newacronym{5g}{5G}{5th generation}
\newacronym{aimd}{AIMD}{Additive Increase Multiplicative Decrease}
\newacronym{am}{AM}{Acknowledged Mode}
\newacronym{amc}{AMC}{Adaptive Modulation and Coding}
\newacronym{aoa}{AoA}{Angle of Arrival}
\newacronym{aod}{AoD}{Angle of Departure}
\newacronym{aqm}{AQM}{Active Queue Management}
\newacronym{awgn}{AGWN}{Additive White Gaussian Noise}
\newacronym{balia}{BALIA}{Balanced Link Adaptation}
\newacronym{bdp}{BDP}{Bandwidth-Delay Product}
\newacronym{bf}{BF}{Beamforming}
\newacronym{cc}{CC}{Congestion Control}
\newacronym{cdf}{CDF}{Cumulative Distribution Function}
\newacronym{cn}{CN}{Core Network}
\newacronym{cnn}{CNN}{convolutional neural network}
\newacronym{cnns}{CNNs}{convolutional neural networks}
\newacronym{cqi}{CQI}{Channel Quality Information}
\newacronym{cp}{CP}{Control Plane}
\newacronym{csirs}{CSI-RS}{Channel State Information - Reference Signal}
\newacronym{wsc}{WSC}{wireless signal classification}
\newacronym{dc}{DC}{Dual Connectivity}
\newacronym{dce}{DCE}{Direct Code Execution}
\newacronym{dci}{DCI}{Downlink Control Information}
\newacronym{dmr}{DMR}{Deadline Miss Ratio}
\newacronym{dmrs}{DMRS}{DeModulation Reference Signal}
\newacronym{e2e}{E2E}{End-to-End}
\newacronym{ecn}{ECN}{Explicit Congestion Notification}
\newacronym{ebs}{EBS}{exhaustive beam sweep}
\newacronym{edf}{EDF}{Earliest Deadline First}
\newacronym{enb}{eNB}{evolved Node Base}
\newacronym{epc}{EPC}{Evolved Packet Core}
\newacronym{es}{ES}{Edge Server}
\newacronym{fdma}{FDMA}{Frequency Division Multiple Access}
\newacronym{fdd}{FDD}{Frequency Division Duplexing}
\newacronym[firstplural=Radio Access Technologies (RATs)]{rat}{RAT}{Radio Access Technology}
\newacronym{fs}{FS}{Fast Switching}
\newacronym{txer}{TX}{transmitter}
\newacronym{rxer}{RX}{receiver}
\newacronym{bt}{BT}{beam tracking}
\newacronym{ftp}{FTP}{File Transfer Protocol}
\newacronym{gnb}{gNB}{Next Generation Node Base}
\newacronym{bs}{BS}{Base Station}
\newacronym{harq}{HARQ}{Hybrid Automatic Repeat reQuest}
\newacronym{hetnet}{HetNet}{Heterogeneous Network}
\newacronym{hh}{HH}{Hard Handover}
\newacronym{hol}{HOL}{Head-of-Line}
\newacronym{ber}{BER}{bit error rate}
\newacronym{dsp}{DSP}{digital signal processing}
\newacronym{ia}{IA}{initial access}
\newacronym{imt}{IMT}{International Mobile Telecommunication}
\newacronym{iot}{IoT}{Internet of Things}
\newacronym{los}{LOS}{Line-of-Sight}
\newacronym{lte}{LTE}{Long Term Evolution}
\newacronym{m2m}{M2M}{Machine to Machine}
\newacronym{ml}{ML}{Machine Learning}
\newacronym{dl}{DL}{deep learning}
\newacronym{fpga}{FPGA}{field-programmable gate array}
\newacronym{drl}{DRL}{Deep Reinforcement Learning}
\newacronym{mac}{MAC}{Medium Access Control}
\newacronym{mc}{MC}{Multi-Connectivity}
\newacronym{mcs}{MCS}{Modulation and Coding Scheme}
\newacronym{mec}{MEC}{Mobile Edge Cloud}
\newacronym{mi}{MI}{Mutual Information}
\newacronym{mimo}{MIMO}{Multiple Input, Multiple Output}
\newacronym{mmwave}{mmWave}{millimeter wave}
\newacronym{mmWave}{mmWave}{Millimeter wave}
\newacronym{mptcp}{MPTCP}{Multipath TCP}
\newacronym{mr}{MR}{Maximum Rate}
\newacronym{mss}{MSS}{Maximum Segment Size}
\newacronym{mtd}{MTD}{Machine-Type Device}
\newacronym{mtu}{MTU}{Maximum Transmission Unit}
\newacronym{nfv}{NFV}{Network Function Virtualization}
\newacronym{nlos}{NLOS}{Non-Line-of-Sight}
\newacronym{nr}{NR}{New Radio}
\newacronym{ofdm}{OFDM}{Orthogonal Frequency Division Multiplexing}
\newacronym{pdcch}{PDCCH}{Physical Downlink Control Channel}
\newacronym{pdcp}{PDCP}{Packet Data Convergence Protocol}
\newacronym{pdsch}{PDSCH}{Physical Downlink Shared Channel}
\newacronym{pdu}{PDU}{Packet Data Unit}
\newacronym{pf}{PF}{Proportional Fair}
\newacronym{pgw}{PGW}{Packet Gateway}
\newacronym{phy}{PHY}{Physical}
\newacronym{pbch}{PBCH}{Physical Broadcast Channel}
\newacronym[plural=\gls{mme}s,firstplural=Mobility Management Entities (MMEs)]{mme}{MME}{Mobility Management Entity}
\newacronym[plural=\gls{fir}s,firstplural=Finite Input Response (MMEs) ]{fir}{FIR}{Finite Impulse Response}
\newacronym{prb}{PRB}{Physical Resource Block}
\newacronym{pss}{PSS}{Primary Synchronization Signal}
\newacronym{pucch}{PUCCH}{Physical Uplink Control Channel}
\newacronym{pusch}{PUSCH}{Physical Uplink Shared Channel}
\newacronym{rach}{RACH}{Random Access Channel}
\newacronym{ran}{RAN}{Radio Access Network}
\newacronym{red}{RED}{Random Early Detection}
\newacronym{rf}{RF}{Radio Frequency}
\newacronym{rlc}{RLC}{Radio Link Control}
\newacronym{rlf}{RLF}{Radio Link Failure}
\newacronym{rrc}{RRC}{Radio Resource Control}
\newacronym{rrm}{RRM}{Radio Resource Management}
\newacronym{rr}{RR}{Round Robin}
\newacronym{rs}{RS}{Remote Server}
\newacronym{rsrp}{RSRP}{Reference Signal Received Power}
\newacronym{rss}{RSS}{Received Signal Strength}
\newacronym{rtt}{RTT}{Round Trip Time}
\newacronym{rw}{RW}{Receive Window}
\newacronym{rx}{RX}{Receiver}
\newacronym{sa}{SA}{standalone}
\newacronym{sack}{SACK}{Selective Acknowledgment}
\newacronym{sap}{SAP}{Service Access Point}
\newacronym{ap}{AP}{Access Point}
\newacronym{sch}{SCH}{Secondary Cell Handover}
\newacronym{scoot}{SCOOT}{Split Cycle Offset Optimization Technique}
\newacronym{sdma}{SDMA}{Spatial Division Multiple Access}
\newacronym{sinr}{SINR}{Signal to Interference plus Noise Ratio}
\newacronym{sm}{SM}{Saturation Mode}
\newacronym{snr}{SNR}{Signal-to-Noise-Ratio}
\newacronym{son}{SON}{Self-Organizing Network}
\newacronym{ss}{SS}{Synchronization Signal}
\newacronym{ssbs}{SSBs}{synchronization signal blocks}
\newacronym{ssb}{SSB}{synchronization signal block}
\newacronym{srs}{SRS}{Sounding Reference Signal}
\newacronym{sss}{SSS}{Secondary Synchronization Signal}
\newacronym{tb}{TB}{Transport Block}
\newacronym{tcp}{TCP}{Transmission Control Protocol}
\newacronym{tdd}{TDD}{Time Division Duplexing}
\newacronym{tdma}{TDMA}{Time Division Multiple Access}
\newacronym{tfl}{TfL}{Transport for London}
\newacronym{tm}{TM}{Transparent Mode}
\newacronym{trp}{TRP}{Transmitter Receiver Pair}
\newacronym{tti}{TTI}{Transmission Time Interval}
\newacronym{ttt}{TTT}{Time-to-Trigger}
\newacronym{tx}{TX}{Transmitter}
\newacronym{ue}{UE}{User Equipment}
\newacronym{ul}{UL}{Uplink}
\newacronym{uml}{UML}{Unified Modeling Language}
\newacronym{um}{UM}{Unacknowledged Mode}
\newacronym{utc}{UTC}{Urban Traffic Control}
\newacronym{vm}{VM}{Virtual Machine}
\newacronym{rsrq}{RSRQ}{Reference Signal Received Quality}
\newacronym{rssi}{RSSI}{Received Signal Strength Indicator}
\newacronym{crs}{CRS}{Cell Reference Signal}
\newacronym{nsa}{NSA}{Non Stand Alone}
\newacronym{mrdc}{MR-DC}{Multi \gls{rat} \gls{dc}}
\newacronym{endc}{EN-DC}{E-UTRAN-\gls{nr} \gls{dc}}
\newacronym{5gc}{5GC}{5G Core}
\newacronym{si}{SI}{Study Item}
\newacronym{iab}{IAB}{Integrated Access and Backhaul}
\newacronym{wf}{WF}{Wired-first}
\newacronym{hqf}{HQF}{Highest-quality-first}
\newacronym{pa}{PA}{Position-aware}
\newacronym{mlr}{MLR}{Maximum-local-rate}
\newacronym{wbf}{WBF}{Wired Bias Function}
\newacronym{mib}{MIB}{Master Information Block}
\newacronym{sib}{SIB}{Secondary Information Block}
\newacronym{kpi}{KPI}{Key Performance Indicator}
\newacronym{ppp}{PPP}{Poisson Point Process}
\newacronym{gtp}{GTP}{GPRS Tunneling Protocol}
\newacronym{amf}{AMF}{Access and Mobility Management Function}
\newacronym{dash}{DASH}{Dynamic Adaptive Streaming over HTTP}
\newacronym{http}{HTTP}{HyperText Transfer Protocol}
\newacronym{qos}{QoS}{Quality of Service}
\newacronym{udp}{UDP}{User Datagram Protocol}
\newacronym{cu}{CU}{Central Unit}
\newacronym{du}{DU}{Distributed Unit}
\newacronym{mt}{MT}{Mobile Termination}
\newacronym{sdap}{SDAP}{Service Data Adaptation Protocol}
\newacronym{tdm}{TDM}{Time Division Multiplexing}
\newacronym{fdm}{FDM}{Frequency Division Multiplexing}
\newacronym{sdm}{SDM}{Space Division Multiplexing}
\newacronym{dag}{DAG}{Directed Acyclic Graph}
\newacronym{st}{ST}{Spanning Tree}
\newacronym{ummimo}{UM-MIMO}{Ultra-massive Multiple Input, Multiple Output}
\newacronym{uavs}{UAVs}{Unmanned Aerial Vehicles}
\newacronym{wlan}{WLAN}{Wireless LAN}
\newacronym{rlnc}{RLNC}{Random Linear Network Coding}
\newacronym{drx}{DRX}{Discontinuous Reception}
\newacronym{cpu}{CPU}{Central Processing Unit}
\newacronym{txb}{TXB}{transmitter's beam}
\newacronym{rxb}{RXB}{receiver's beam}
\makeglossaries
\glsresetall

\begin{document}

\maketitle
\pagenumbering{gobble}

\begin{abstract}
Due to the sheer scale of the \gls{iot} and 5G, the wireless spectrum is becoming severely congested. For this reason, wireless devices will need to continuously adapt to current spectrum conditions by changing their communication parameters in real-time. Therefore, \gls{wsc} will become a compelling necessity to decode fast-changing signals from dynamic transmitters. Thanks to its capability of classifying complex phenomena without explicit mathematical modeling, \gls{dl} has been demonstrated to be a key enabler of \gls{wsc}. Although \gls{dl} can achieve a very high accuracy under certain conditions, recent research has unveiled that the wireless channel can disrupt the features learned by the \gls{dl} model during training, thus drastically reducing the classification performance in real-world live settings. Since retraining classifiers is cumbersome after deployment, existing work has leveraged the usage of carefully-tailored \gls{fir} filters that, when applied at the transmitter's side, can restore the features that are lost because of the the channel actions, i.e., \textit{waveform synthesis}. However, these approaches compute \glspl{fir} using offline optimization strategies, which limits their efficacy in highly-dynamic channel settings. In this paper, we improve the state of the art by proposing Chares, a \gls{drl}-based framework for channel-resilient adaptive waveform synthesis. Chares adapts to \textit{new and unseen} channel conditions by optimally computing through \gls{drl} the \glspl{fir} in real time. Chares is a DRL agent whose architecture is based upon the Twin Delayed Deep Deterministic Policy Gradients (TD3), which requires minimal feedback from the receiver and explores a continuous action space for best performance. Chares has been extensively evaluated on two well-known datasets with an extensive number of channels. We have also evaluated the real-time latency of Chares with an implementation on \gls{fpga}. Results show that Chares increases the accuracy up to 4.1x when no waveform synthesis is performed, by 1.9x with respect to existing work, and can compute new actions within 41$\mu$s.\vspace{-0.7cm}
\end{abstract}


\glsresetall

\section{Introduction} \label{sec:intro}

The rise of the \gls{iot} and \gls{5g} networks will mark an era where several billion people and devices will ubiquitously request services using a multitude of networking protocols and architectures \cite{yaqoob2017internet}. The inevitable outcome  will be an extremely crowded spectrum (especially in the sub-6GHz regime) where diverse technologies coexist and share the same spectrum bands. To solve this problem -- known as \textit{spectrum crunch} -- the networking community is undertaking a radical paradigm shift where inflexible architectures are being left behind in favor of ``smart'' transceivers that utilize spectrum resources more efficiently by reconfiguring networking parameters and transmission strategies in real time. Dynamic spectrum access \cite{jin2018specguard,chiwewe2017fast,agarwal2016edsa}, spectrum sensing \cite{shokri2016spectrum,vazquez2018hybrid,lv2018cognitive,zheleva2018airview}, and reconfigurable transceivers \cite{restuccia2019big,restuccia2020deepwierl} are just a few examples of technologies that will become compelling necessities.

Being able to classify phenomena without explicit modeling, \gls{dl} architectures \cite{lecun2015deep}, and in particular \gls{cnns}, have experienced a surge of interest from the community over the last few years as flexible and efficient tools to perform a plethora of networking-related tasks -- such as modulation classification \cite{Kulin-ieeeaccess2018,deepsig,Karra-ieeedyspan2017,o2017introduction,xiong2019robust}, radio fingerprinting \cite{zheng2019fid,16-Peng-ieeeiotj2018,17-Xie-ieeeiotj2018,18-Xing-ieeecomlet2018}, and real-time radio-frequency (RF) front-end reconfiguration \cite{restuccia2019big,restuccia2020deepwierl}. For this reason, \gls{dl} systems are among the most promising tools to develop and deploy the next-generation of intelligent and adaptive wireless networks \cite{Mao-ieeecomm2018}.

\begin{figure}[b]
    \centering
    \includegraphics[width=0.85\columnwidth]{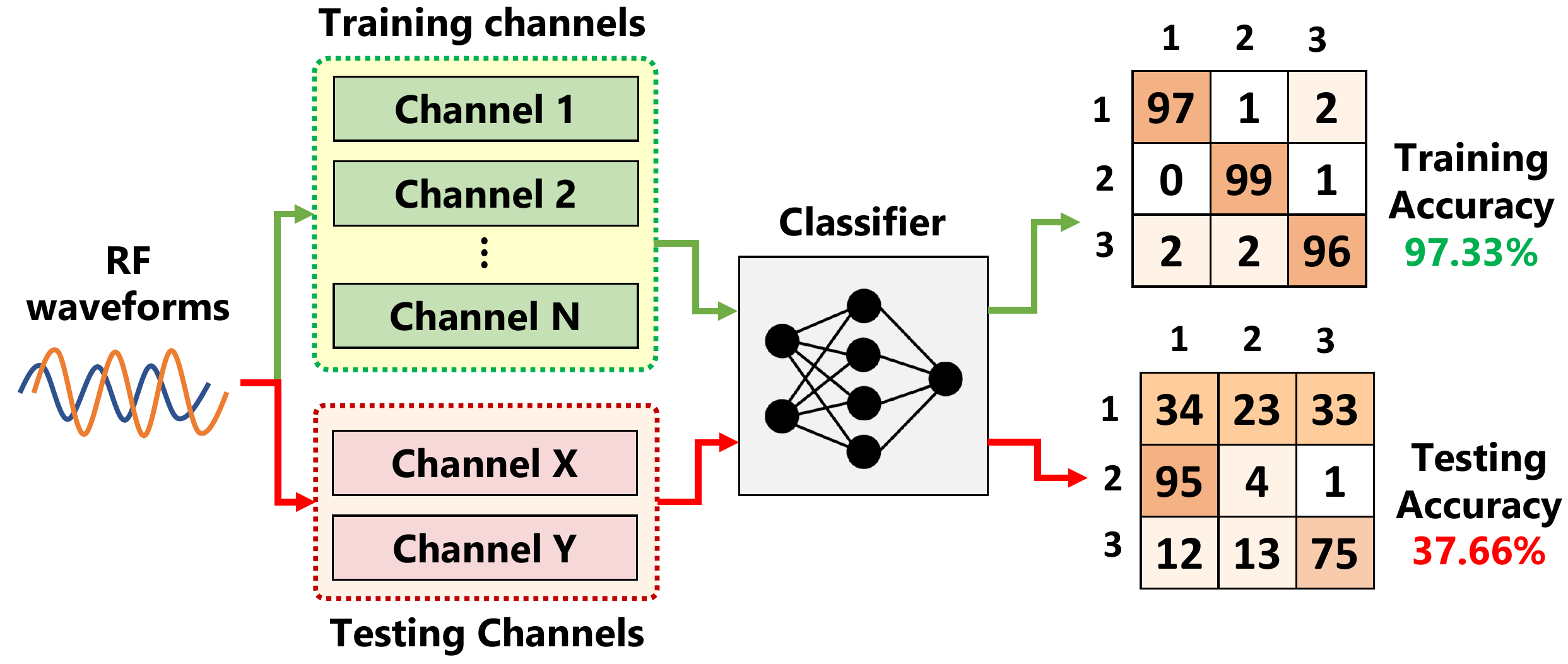}
    \vspace{-0.2cm}
    \caption{Impact of different channel conditions on modulation classification.}
    \label{fig:intro}
 \end{figure}

The majority of the above solutions operate in the complex-domain and can be generalized as \gls{wsc} problems, where waveforms coming from the RF front-end are fed to, and classified by, a neural network \cite{sankhe2019oracle,riyaz2018deep,chen2019artificial,restuccia2019big,restuccia2020deepwierl}. Most of the existing research has focused on demonstrating the effectiveness of \gls{wsc} in classifying different spectrum phenomena. For example, O'Shea \emph{et al.} \cite{deepsig} have demonstrated that on the 24-modulation dataset considered, \gls{dl} models achieve on the average about 20\% higher classification accuracy  than legacy learning models. Moreover, Restuccia \emph{et al.} \cite{restuccia2019deepradioid} have proven that \gls{dl} can achieve 27\% improvement over traditional \gls{ml} in large-scale radio fingerprinting. However, research has unveiled that the time-varying nature of the wireless channel (\textit{e.g.}, fading, mobility, interference) could have detrimental consequences to the accuracy of the model when tested with samples collected in different times than when the model was trained  \cite{restuccia2019deepradioid,shawabka2020exposing}.

Figure \ref{fig:intro} provides an illustration of the problem at hand. Since the classifier cannot be trained on all possible channel distributions (and realizations), the accuracy will necessarily drop with channels (\textit{e.g.}, $X,Y$) that are different from the ones seen by the classifier during training (\textit{e.g.}, $1,2,\dots,N$). It has been shown that in some cases, the accuracy can drop close to random guessing \cite{shawabka2020exposing}, thus making the classifier unusable and unreliable. At the same time, although one might think to
re-train or fine-tune the model with new data, once deployed \gls{dl} systems implemented in hardware are difficult to retrain in real time as devices usually have limited computational capabilities, and storage might not be always enough to maintain a training dataset. In other words, without any intervention, all the time (hours/days) spent training a model would be wasted.

Existing work \cite{restuccia2019deepradioid} has proposed the use of optimized \gls{fir} filters at the transmitter's side which -- by manipulating the position in the complex space of transmitted IQ samples -- "\textit{synthesize}" waveforms such that the salient features of each class are preserved at the receiver's side despite the negative effect of channel action without compromising the \gls{ber}. Waveform synthesis is then performed by using \glspl{fir} that are computed offline using traditional mathematical optimization. The authors show how this approach can boost classification accuracy by 58\% under different channel conditions. However, the core issue is that \glspl{fir} are static and used in a deterministic fashion by applying them to all transmissions without considering the current channel conditions. In Section \ref{sec:numerical}, we show that such an approach underperforms under diverse and highly-dynamic channel conditions. Moreover, existing work does not take real-time aspects into consideration. Since channels usually change in a matter of milliseconds, it is imperative to design techniques fast enough to operate within channel coherence time. \smallskip

\textbf{Summary of Main Contributions}.~
This paper presents Chares, a \textit{real-time} and \textit{model-free} \gls{drl} approach that leverages waveform synthesis to improve the resilience of WSC applications via channel-specific FIR filtering (Section \ref{sec:proposed}). Beyond playing video games \cite{mnih2013playing}, \gls{drl} has experienced a surge of interest in the wireless networking community as well \cite{Feng-ieeecommag2019,Zhang-infocom2019,Yu-jsac2019,Zhang-ieeetwc2019,Pang-infocom2019,Wang-infocom2019,Zhang2-infocom2019,Wang-infocom2020,wang2020intelligent}. This is because \gls{drl} provides a very general framework based on partially-observable Markov decision process (POMDP), which allows to dynamically solve a multitude of problems without explicit modeling. The key advantage of \gls{drl} is the fact that the model is trained in an online fashion, and therefore it can be fine-tuned according to the current channel conditions \cite{restuccia2020deepwierl}. Chares has been designed to be independent of the specific \gls{wsc} problem, does not require gradients and/or channel models, and only minimal feedback information (\textit{i.e.}, prediction label and/or output of the classifier) is required to synthesize waveforms. To achieve channel resilience, Chares \gls{drl} agent is based upon the Twin Delayed Deep Deterministic Policy Gradients (TD3) DRL framework (Section \ref{sec:arch_feed}) and it is trained off-policy by voluntarily perturbing its computed actions, \textit{i.e.}, the FIR filters, with random Gaussian noise. This procedure allows the agent to learn policies that are robust against noise and perturbation.

We assess the performance of Chares for two different \gls{wsc} problems, namely modulation classification and radio fingerprinting and show that Chares can be transferred across the two problems without any modification (Section \ref{sec:numerical}). We show that Chares always improve the accuracy of the classifier with gains up to 4.1x if compared to baseline solutions and up to 1.9x if compared to the state-of-the-art \cite{restuccia2019deepradioid} where FIR filters are computed offline with white-box approaches. Furthermore, we show that while other solutions do not adapt to unseen channel conditions, Chares can effectively increase the accuracy of the WSC system in both single- and multi-label problems even in the case of adversarial attacks such as jammers. Finally, to demonstrate that Chares can fully operate within typical channel coherence times, we have performed an \gls{fpga} implementation of the state-action policy function of Chares. We show that its latency is below 41$\mu$s, orders of  magnitude less than coherence time with mobile users in WiFi applications, which is approximately 40ms \cite{xie2018precise}.\vspace{-0.2cm}



\glsresetall

\section{Related Work} \label{sec:related}

Thanks to their ability to model complex classification and optimization problems, \gls{dl} is enjoying tremendous popularity. Some applications include modulation classification \cite{Kulin-ieeeaccess2018,deepsig,Karra-ieeedyspan2017,o2017introduction,xiong2019robust}, radio fingerprinting \cite{zheng2019fid,16-Peng-ieeeiotj2018,17-Xie-ieeeiotj2018,18-Xing-ieeecomlet2018}, and real-time radio-frequency (RF) front-end reconfiguration \cite{restuccia2019big,restuccia2020deepwierl}. We refer the reader to \cite{luong2018applications,jagannath2019machine} for exhaustive surveys on the topic.

Despite its strengths, it has been recently demonstrated that \gls{dl} and its efficacy in \gls{wsc} applications is severely hindered by time-varying channel conditions \cite{shawabka2020exposing}. Similarly, also \gls{drl} has been successfully used in a plethora of different contexts in the networking field. Among others, we mention resource allocation and network virtualization \cite{Feng-ieeecommag2019,He-ieeetvt2018,Sun-ieeeiotj2019,Li-ieeeaccess2018,Zhang-infocom2019},  dynamic spectrum access \cite{Yu-jsac2019,Wang-tccn2018,Naparstek-ieeetwc2019}, cellular networks \cite{Liu-ieeeiotj2018,Wang-ieeeiotj2018},  rate selection \cite{Zhang-ieeetwc2019}, and multimedia streaming \cite{Pang-infocom2019,Wang-infocom2019,Zhang2-infocom2019,Wang-infocom2020,wang2020intelligent}.
Although \gls{drl} has been frequently used for optimization of networking metrics, it has seldom been used so far to improve the performance of other learning systems in the networking context. For example, Wang \textit{et al.} \cite{wang2020federated} use DRL to optimize federated learning of non-IID data. Zhan \emph {et al.} \cite{zhan2020incentive} design a \gls{drl}-based incentive mechanism for efficient edge learning. To the best of our knowledge, however, there are no \gls{drl} applications to improve the resiliency of learning systems that are already deployed and whose re-training is neither possible nor efficient.

Different from existing work, in this paper we consider \gls{wsc} problems, which are extremely sensitive to channel fluctuations to the point that they cause the accuracy to drop close to random guessing in real-world wireless environments \cite{shawabka2020exposing}.  Although this problem is extremely relevant to many \gls{wsc} use-cases, only recently has it received attention from the community \cite{restuccia2019deepradioid}. Although the approach in \cite{restuccia2019deepradioid} has proved its effectiveness on RF fingerprinting \gls{wsc} problems, it is  an offline approach that leverages optimization algorithms requiring gradients of the classifier, which makes this approach not tailored to real-time applications requiring online optimization and where gradients might not be available. Our work separates itself from \cite{restuccia2019deepradioid} since we consider an \textit{online} approach that leverages \gls{drl} to counteract channel fluctuations with minimal feedback in real time. Due to the stochastic nature of the channel, we leverage off-policy training to voluntarily add noise to learned policies which makes Chares resilient against channel fluctuations and unseen conditions. We demonstrate that our approach can operate within the coherence time of the wireless channel. We show that it always outperforms the state-of-the-art by achieving up to 4.1$\times$ improvement in terms of classification accuracy, and how Chares can be effectively used for single- and multi-label WSC problems even in the case of adversarial attacks.
\vspace{-0.1cm}


\section{Our Approach: DRL-based waveform synthesis} \label{sec:proposed}

We now present a radically new approach called Chares\footnote{The name comes from the ancient Greek builder of the Colossus of Rhodes.}, a \gls{drl}-based adaptive system for channel-resilient WSC applications. Rather than re-training the classifier, Chares adds carefully crafted distortions to the transmitted waveform, aiming at \textit{restoring} and \textit{amplifying} signal features that are lost after distortions introduced by the wireless channel. Our approach is especially helpful in cases where (i) data is scarce and unbalanced; and (ii) the node has limited computational resources to rapidly re-train and fine-tune the model. As we will describe soon, this is achieved by leveraging \gls{fir} filtering. Since different channel conditions affect received waveforms -- and their features -- differently, Chares distorts transmitted waveforms on a per-channel basis by generating FIR filters that are hand-tailored to each channel condition. Before going into the details on Chares architecture and operations, in the following we provide a brief overview on FIR filtering and how it can be used to artificially distort transmitted waveforms to increase classification accuracy.

\vspace{0.01cm}
\textbf{Fundamentals of FIRs.~}FIR filtering is a DSP technique that makes it possible to filter signals via a limited number of coefficients, \textit{i.e.}, the \textit{FIR taps}. Although FIR filters are usually employed to suppress undesired frequency components of wireless signals, it is possible to change the amplitude and phase of transmitted waveforms in the complex plane, \textit{i.e.}, introduce artificial distortions to the signal,  by properly tuning the values of each FIR tap. In other words, for any given complex-valued signal $\mathbf{x}=(x[n])_{n=1,\dots,N}$, and FIR filter with complex taps $\mathbf{h}=(h[m])_{m=1,\dots,M}$, the $n$-th filtered element of $\mathbf{x}$ can be expressed as follows: \vspace{-0.3cm}

\begin{equation} \label{eq:FIR:general}
    \tilde{x}[n] = \sum_{m=0}^{M-1} h[m] x[n-m]
    \vspace{-0.2cm}
\end{equation}

The advantages of FIR filtering for wireless applications are manifold: (i) FIRs have a linear input/output relationship that can be expressed in closed-form; (ii) the output can be computed via fast discrete convolution algorithms in $\mathcal{O}(N\log N)$, thus allowing their usage for real-time applications; (iii) FIRs can efficiently control the position of transmitted IQ samples in the complex-space with just a few taps; and (iv) FIRs can be compensated out from the received waveform at the receiver side, thus removing any distortion added by the the FIR\footnote{Any convolution $x \ast h = y $ between two signals $x,h$ in the time domain translates into a multiplication $X(f) H(f) = Y(f)$ in the frequency domain. This relationship can be utilized to reconstruct the signal $x$ through the relationship $X(f)=Y(f)/H(f)$ and an inverse Fourier transform.}.

\begin{figure}[t]
    \centering
    \includegraphics[width=0.8\columnwidth]{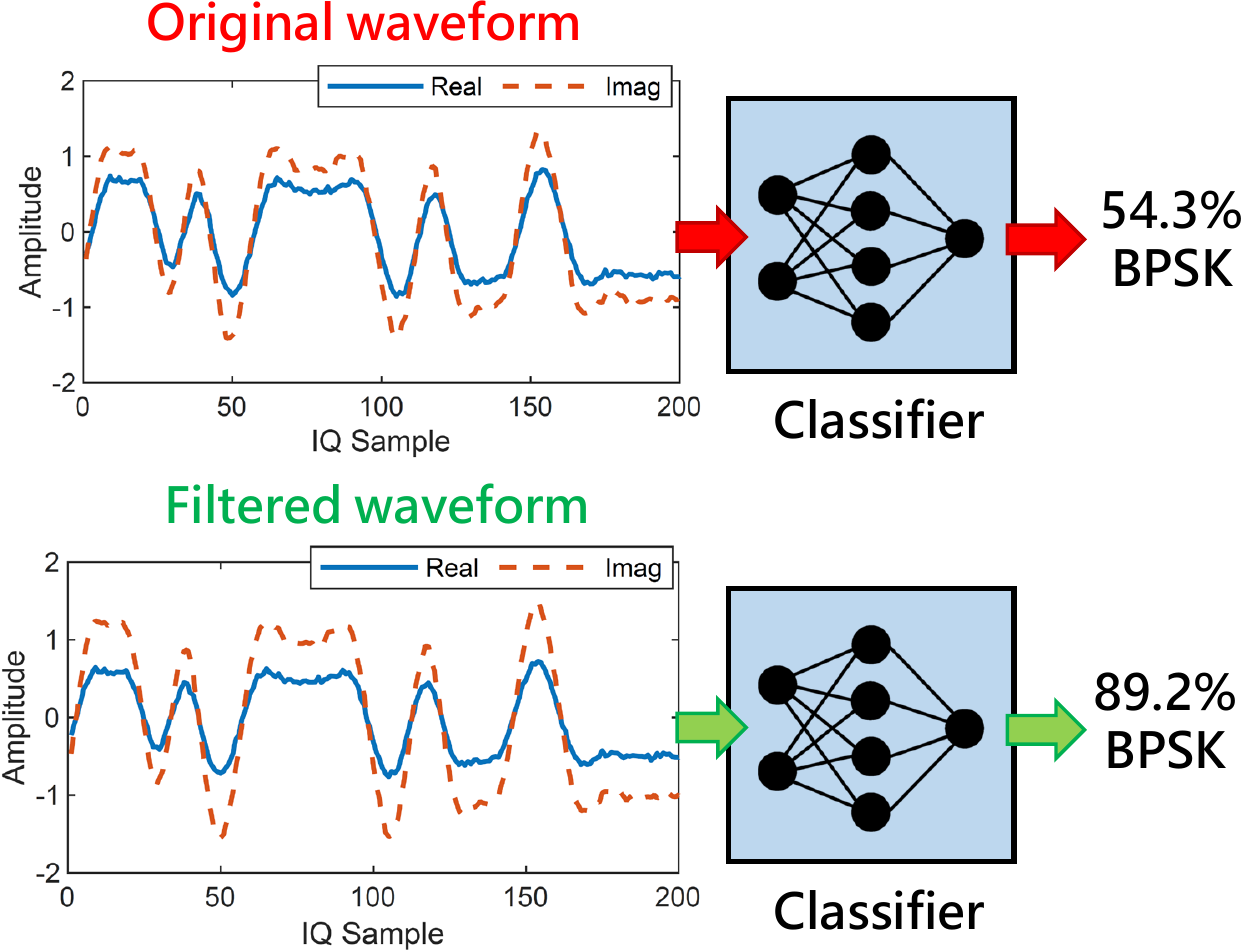}
    \vspace{-0.2cm}
    \caption{Classification accuracy of a BPSK modulated signals with (bottom) and without (top) FIR filtering.}
    \label{fig:example_mod}
 \end{figure}

Although details on the methodology to compute effective FIR filters for channel-resilient WSC will be given in the following sections, here we provide an illustrative example demonstrating the potential of FIR filtering for ML-based wireless communications. We have trained a modulation-recognition NN classifier on the DeepSig RADIOML 2018.01A dataset (\url{https://www.deepsig.ai/datasets}) -- details on the NN and the dataset are reported in Section \ref{sec:numerical:datasets}). In \reffig{fig:example_mod}, we show two BPSK waveforms as well as the output of the classifier specifying the probability that each waveform is classified as BPSK. The first waveform (top) has been extracted from the dataset and fed to the classifier, with a prediction accuracy of 54.3\%. The second one has been obtained by filtering the same waveform with a FIR filter that has been computed with the approach presented in this paper. As clearly shown by \reffig{fig:example_mod}, FIR filtering increased classification accuracy by 64\%.

 \vspace{-0.2cm}
\section{Chares Architecture} \label{sec:arch_feed}

\begin{figure*}[t!]
    \centering
    \includegraphics[width=0.8\textwidth]{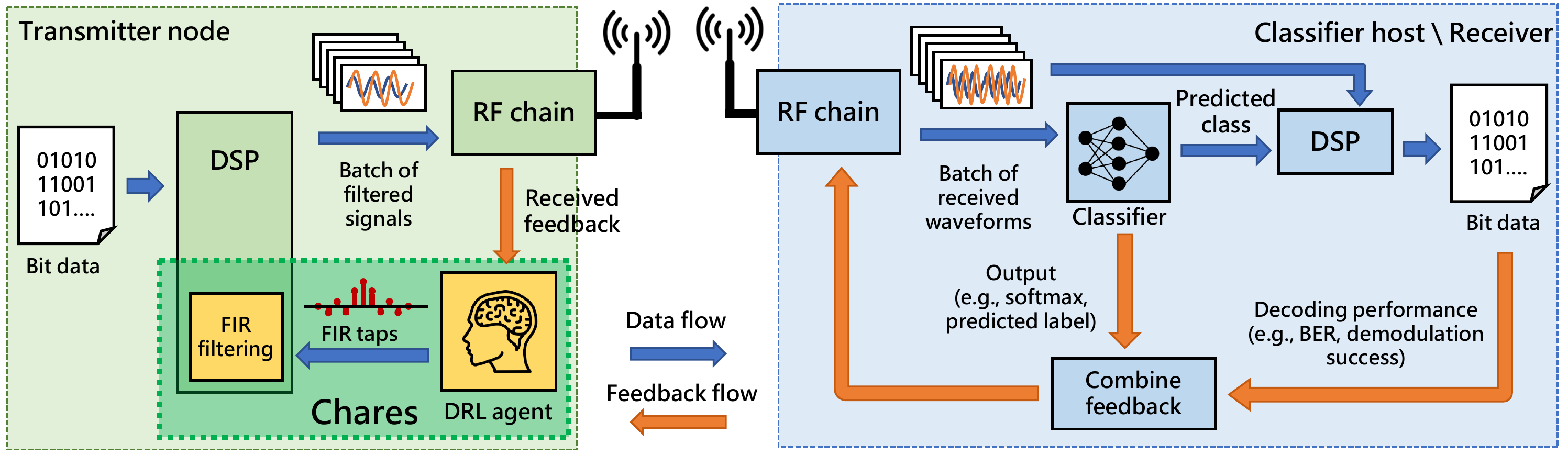}
    \vspace{-0.2cm}
    \caption{The considered system model and Chares building blocks.}
    \label{fig:model}
 \end{figure*}

\textbf{System Model.~}We consider the communication system in \reffig{fig:model}. We assume the receiver is equipped with a wireless signal classifier (\textit{e.g.}, a neural network) trained to perform \gls{wsc} tasks such as radio fingerprinting, modulation recognition, and so on. We do not make any assumption on the specific  classifier's structure (which hereafter is referred to as the \textit{classifier}), and  assume that the classifier outputs a single label identifying the predicted class out of $C$ possible classes. For example, in the case of modulation classification, the classifier is fed with received waveforms and outputs the predicted modulation scheme that has been applied to those waveforms (\textit{e.g.}, BPSK, QPSK, etc). As shown in \reffig{fig:model}, we also assume the output of the classifier is actively utilized within the receive chain to further process the received waveforms. For example, in the case of modulation recognition the predicted labels is used to demodulate and decode received signals, while in RF fingerprinting applications waveforms might be decoded and deciphered by using transmitted-specific encryption keys. We point out that in our system, the  accuracy of the classifier plays a vital role in the communication process, as  misclassifications would inevitably results in decoding errors.

\textbf{Chares Architecture.}~The architecture of Chares as well as the main procedures of the system are illustrated in \reffig{fig:model}. Chares operates at the transmitter's side only and consists of two main components: the DRL agent and the FIR filtering component. We consider a \textit{black-box} approach where the transmitter has no access to the classifier (\textit{e.g.}, model, weights) and can only receive \textit{partial feedback} from the receiver hosting the classifier periodically. Specifically, we assume that Chares applies a specific set of FIR taps to a set of consecutive waveforms, and the receiver feeds back relevant information regarding these waveforms to the transmitter. We consider the case where receiver and transmitter cooperate to improve the accuracy of the WSC task, hence the feedback generated by the receiver is truthful. As shown in \reffig{fig:model}, we envision two potential classes of feedback:
\begin{itemize}
    \item \textit{Classifier-specific}: this class includes any information available at the output of the classifier such as predicted labels (out of $C$ possible classes) and softmax output of the classifier. For the sake of generality, we consider both the case where the receiver can either feed back the above metrics for all the $W$ waveforms in the batch, or only send average results across the whole batch. In Section \ref{sec:numerical}, we will show that average values are enough for Chares to increase the classification accuracy;
    \item \textit{Communication-specific}: such as BER, percentage of successfully decoded packets and demodulated signals.
\end{itemize}

To better understand the importance of this feedback, let us assume that the receiver uses a neural network (NN) classifier to first recognize the modulation of incoming waveforms and demodulate them by using NN inference. If the classification procedure is correct, the receiver would be able to properly demodulate and decode received data. On the contrary, in the case of misclassification -- due to channel impairments -- the receiver would not be able to demodulate received signals, eventually resulting in higher demodulation errors. \smallskip

\textbf{Chares Operations.}~After deployment of transmitter and receiver, we assume that transmitter's data  is processed \textit{as is} by DSP units and converted into waveforms that are transmitted over the wireless channel. Upon reception, the receiver extracts $W>0$ IQ samples sequences (\textit{i.e.}, a \textit{batch}) of length $L$, where $L$ represents the input size of the classifier. The latter processes the extracted batch and outputs the probability that the input belongs to a specific class (\textit{e.g.}, modulation or device identifier in the case of modulation recognition or RF fingerprinting, respectively) as well as the final predicted class. Then, received waveforms are fed to the DSP module that leverages the output of the classifier to properly decode received signals.

Once received waveforms are processed, the receiver generates feedback containing prediction and decoding results that are sent back to the transmitter. Upon reception of such feedback, Chares DRL agent decides whether or not to compute a new set of FIR taps to be applied to future transmissions so as to improve the classification accuracy of the classifier (details on this procedure will be given in Section \ref{sec:architecture}).
For example, if a batch of waveforms has been successfully demodulated by the receiver, the DRL agent does not need to compute new FIR taps as transmitted waveforms are already being classified correctly. On the contrary, if channel action is introducing distortions that result in the receiver reporting the wrong classification labels, Chares DRL agent reacts to counteract channel action and computes a new set of FIR taps that would restore signal features and increase the accuracy of the classifier.


To summarize, Chares brings the following major and fundamental advantages to WSC problems:

    $\bullet$ \textit{Model-free:} existing approaches are mostly "white-box", \textit{i.e.}, needing either access to the classifier and its gradients \cite{restuccia2019deepradioid}.
    On the contrary, Chares is black-box, \textit{i.e.}, agnostic   to  channel conditions, wireless model and classification problem. \textit{As mentioned before, the DRL feedback does not embed any information on the specific WSC problem solved by the classifier}. Therefore, Chares is capable of operating in any channel condition and can be applied to a variety of classification problems in the wireless domain. This aspect will be demonstrated in Section \ref{sec:numerical}, where we  show how the same instance of the Chares DRL agent can be applied to RF fingerprinting and modulation classification problems without \textit{any} modification;

    $\bullet$ \textit{Minimal feedback:} the DRL agent must be able to learn how to compute FIR taps by leveraging minimal amount of information from the receiver. Too much information might generate too high overhead, which would eventually result in increased power consumption;

  $\bullet$ \textit{Feature-free design:} Being model-free, Chares does not need to learn the features of the classifier directly. Instead, the Chares DRL agent learns how to select specific actions (e.g., FIR taps) to respond to specific outputs of the classifier. In fact, Chares leverages classifier's output (e.g., softmax), which provides useful information on the activation of neurons at the last layer of the classifier, and thus allows the agent to compute FIR taps that fire the desired neuron of the classifier. As a consequence, the agent learns by itself what are the features that trigger a specific neuron, and learns how to amplify them while counteracting negative channel effects;

  $\bullet$ \textit{Adaptive:} Chares computes new FIR taps as soon  as the receiver reports misclassifications. In this way, Chares can achieve channel-resilient WSC adaptively by rapidly responding to varying and possibly unseen channel conditions. Conversely, existing approaches compute FIR taps offline over entire training datasets. Such an approach has several drawbacks: (i) since FIRs are computed on large amounts of data, taps are effective on average but are not designed to counteract specific channel conditions, meaning that FIRs might work efficiently for some channel conditions but sonorously fail under different channels \cite{restuccia2019deepradioid}; (ii) computational complexity is high due to the size and diversity of the dataset, which prevents fast convergence; and (iii) statically assigned taps do not properly work (and might be harmful) under unseen channel conditions. \vspace{-0.2cm}

\section{Chares DRL agent design} \label{sec:architecture}

Despite the advantages mentioned in previous sections, DRL cannot be considered as a plug-and-play solution. Indeed, the DRL model must be able to capture the features and requirements of the specific application as well as learn how to adapt promptly to diverse inputs. In short, the selection of the DRL agent will determine its success.
To define a DRL system, we need to specify the \textit{environment} where the \textit{agent} operates, the \textit{state} of the environment that can be observed by the agent and the \textit{actions} the agent can take to respond to each observation of the environment, and the corresponding \textit{reward} that the agent uses to score each action.

 \begin{figure}[t]
    \centering
    \includegraphics[width=0.85\columnwidth]{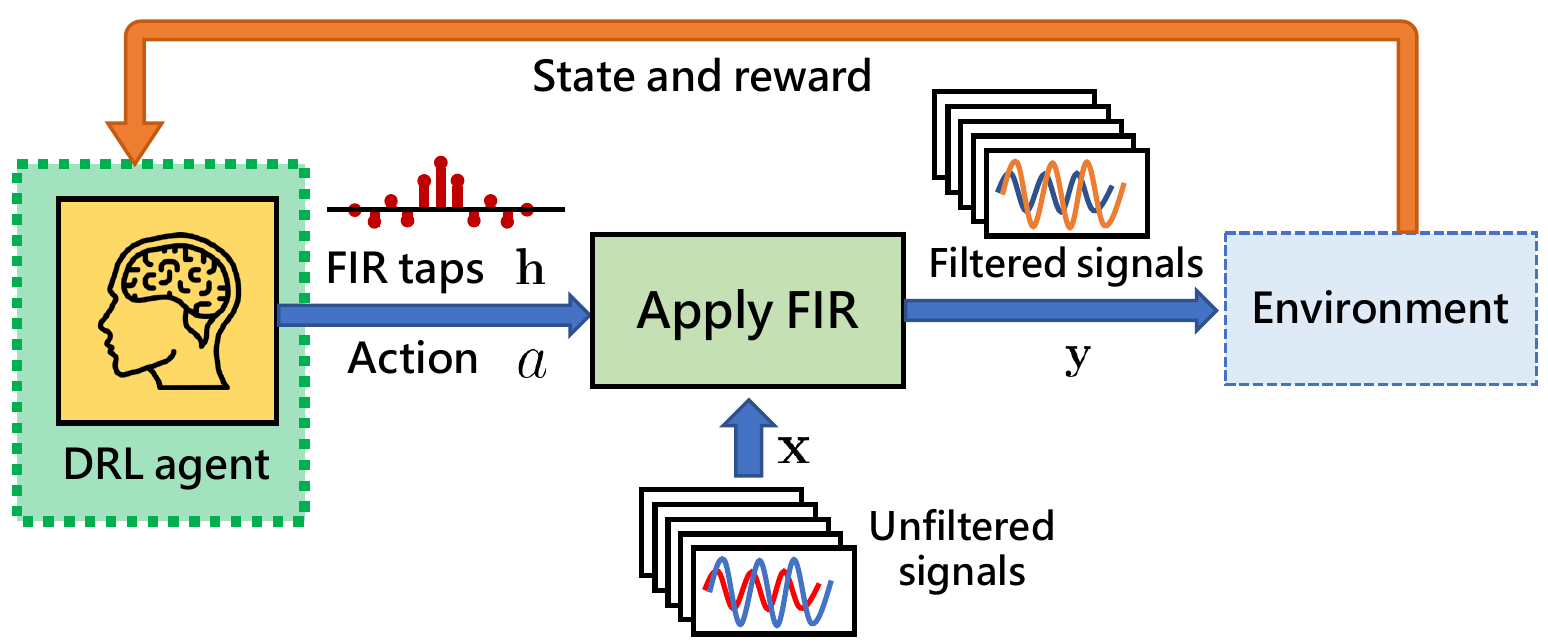}
    \caption{Overview of Chares DRL agent interactions with the environment.}
    \label{fig:drl_overview}
 \end{figure}

Our DRL framework is illustrated as in \reffig{fig:drl_overview}. The environment is identified with the receiver sending feedback (\textit{i.e.}, the observation) on the state $s$ of the classifier (\textit{i.e.}, the accuracy). Similarly, the action $a$ corresponding to the selection of FIR taps $\mathbf{h}$, \textit{i.e.}, $\mathbf{h}=a$, to synthesize waveforms $\mathbf{x}$ according to \eqref{eq:FIR:general} and generate transmitted waveforms $\mathbf{y}$. The reward $r$ -- which we will define formally in Section \ref{sec:reward} -- is then used by the Chares agent to determine whether or not the selected action has improved the accuracy of the classifier at the receiver side.

Let $\states$ be the set of all possible states, and let $\actions$ be the FIR taps space representing the set of actions. We define a default FIR configuration with values $\mathbf{h}^0=[1,0,0,\dots,0]$ representing the case where no distortion is added to the waveform. Also, since waveform synthesis with excessively large FIR taps can potentially distort transmitted waveforms \cite{restuccia2019deepradioid}, we constrain the maximum and minimum values of both real an imaginary parts of each tap. Specifically, let $h[m]$ be the $m$-th tap of a FIR filter $\mathbf{h}$ computed by the Chares DRL agent and $\alpha$ be a small real number.
A feasible FIR filter must satisfy the following conditions $\mathrm{Re}(h[m])\in[\mathrm{Re}(h^0[m])-\alpha, \mathrm{Re}(h^0[m])+\alpha$ and $\mathrm{Im}(h[m])\in[\mathrm{Im}(h^0[m])-\alpha, \mathrm{Im}(h^0[m])+\alpha$, with $h^0[m]\in\mathbf{h}^0$.
FIR taps that satisfy these conditions have been shown to be effective and do not deteriorate transmitted signals and BER significantly \cite{restuccia2019deepradioid}.

As in any DRL problem, we are looking for an agent that learns a policy $\pi(s):\states\rightarrow\actions$, \textit{i.e.}, the \textit{actor policy}, maximizing the discounted sum of rewards $R = \sum_{\tau=0}^T \gamma^{\tau} r_\tau$, where $T$ represents the horizon of the maximization problem and $\gamma>0$ is a term to weigh instantaneous versus future rewards.

Traditional RF algorithms solve the above discounted reward maximization problem via the Bellman's equation. Specifically, they assign a score $Q(s,a) = r + \gamma \max_{a'\in\actions} Q(s',a')$, \textit{i.e.}, the Q-value, to each action-state pair, and compute a policy that selects those actions providing the highest scores. Unfortunately, for high dimensional spaces (such as the one considered in this paper), these approaches result in state-space explosion and are seldom practical.

Before going into the details on how to solve the above problems, we must first consider the following constraints that Chares should also have to satisfy:

$\bullet$ \textit{Noise robustness:} a major challenge in machine learning-based wireless communications is the resiliency of the system against channel noise (e.g., fading, multi-path). Although the DRL agent receives feedback from the receiver, this information is always related to past channel conditions and, although the channel might change slowly over time, the DRL agent must be able to compute FIR taps that are effective against channel conditions that are either completely or slightly different from those experienced by previous transmissions. As a consequence, the designed solution must be robust against noise and stochastic perturbations of channel conditions;

$\bullet$ \textit{Continuous actions:} even though there are many problems where DRL agents are trained to select among a finite set of actions (\textit{e.g.}, move left, pick up an object, select a transmission channel), waveform synthesis relies upon complex-valued FIR filters. This is an extremely relevant feature as minimal variations to taps could distort the waveform to a large extent and negatively impact the classification/decoding process at the receiver side. As a consequence, FIR taps must be continuous so that Chares can achieve fine-grained control over the transmitted waveforms and their IQ samples, thus providing an accurate and reliable tool to synthesize waveforms.\vspace{-0.3cm}



\subsection{Chares DRL architecture} \label{sec:arch:training}

\begin{figure}[t]
    \centering
    \includegraphics[width=0.9\columnwidth]{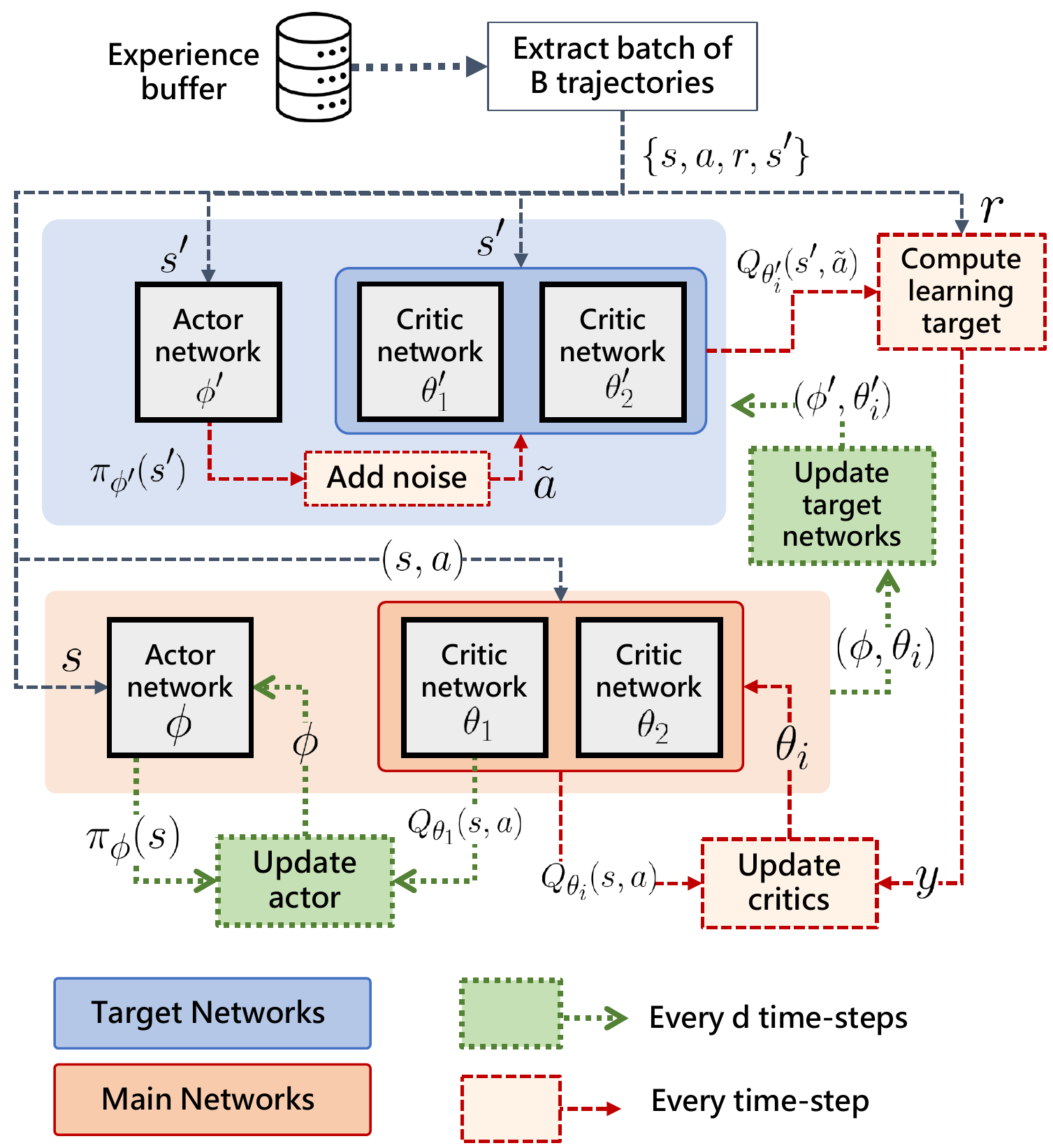}
    \vspace{-0.2cm}
    \caption{DRL agent architecture and training procedures.}
    \label{fig:architecture}
\end{figure}

To address all of the above issues, the Chares DRL agent has been based upon the Twin Delayed Deep Deterministic Policy Gradients (TD3) \cite{td3} model (an extension of the well-established Deep Deterministic Policy Gradients (DDPG) \cite{ddpg} model). \reffig{fig:architecture} provides an overview of the considered architecture whose building blocks and architectural components will be described in detail in the following.

\textbf{Why TD3 as DRL Agent?}~We have selected TD3 because of several reasons. First, it approximates Q-values via deep neural networks (DNNs), thus alleviating the state-space explosion. Specifically, the computation of the optimal policy $\pi(s)$ is   achieved by leveraging an actor-critic setup with (i) one actor network with weights $\phi$ that learns the actor policy $\pi_\phi(s):\states\rightarrow\actions$, and (ii) two critic networks (i.e., the \textit{twins}) with weights $\theta_1$ and $\theta_2$ that are trained to approximate the Q-values $Q_{\theta_i}(s,a)$. At a high-level, the actor is the network that decides with actions to take, and the critics teach the actor how to better select actions. Second, since TD3 supports continuous actions, we can compute gradients of rewards with respect to the learned actor policy. In other words, (i) we can apply deterministic policy gradient \cite{ddpg} to the update learned policies; and (ii) the agent can implement target policy smoothing \cite{td3} where noise is added to actions computed by the target actor policy. This procedure makes the training process less dependent on Q-values, and more robust against estimation errors and noisy wireless channels.

Third, twin critics prevent overestimation of Q-values. Indeed, one-critic  systems are prone to overestimation of the Q-values and thus, biased actor policies \cite{td3}. To overcome this problem, TD3 leverages two critic networks whose weights $\theta_i$ are updated iteratively by "clipping" Q-values estimates of the two critics by considering their minimum only. Fourth, TD3 considers two sets of networks, \textit{i.e.},  the main and target networks, each consisting of one actor and two critics with weights $\phi,\theta_1,\theta_2$ and $\phi',\theta'_1,\theta'_2$, respectively. The main networks are trained at each time-step, and their weights are copied to target networks every $d$ time-steps. This procedure allows to stabilize the training procedure, as main networks are updated with respect to target networks which are frozen for $d$ steps. Fifth, the main actor policy $\pi_\phi$ is updated via a deterministic policy gradient -- however, the update is ``delayed" with respect to the main critic networks. Specifically, the actor policy and target networks are updated every $d$ steps, so that main actor weights $\phi$ are updated through more accurate and stable Q-values.

All of the above features and how to achieve them in practice will be described in detail in the following section. \smallskip

\textbf{Chares Training.~}The training procedure relies upon an experience buffer $\buffer$ storing past experiences of the agent. The $j$-th entry of the buffer is a 4-tuple $(s_j,a_j,r_j,s'_j)$ indicating the action $a_j$ taken by the agent in state $s_j$ which gave a reward $r_j$ and transitioned the environment to state $s'_j$. Since the problem we consider is non-deterministic following the stochastic behavior of the wireless channel, critics cannot compute Q-values directly, which are instead obtained by approximating
\begin{equation} \label{eq:q_val_nondet}
    Q_{\theta_i}(s,a) = r + \gamma \mathbb{E}_{s',a'}\{Q(s',a')\}
\end{equation}
\noindent
where $a'$ is computed via the actor policy $\pi_\phi(s')$ and $s'$ follows an unknown state transition distribution $p(s,a,s')$.

At the beginning of the training, we initialize all DNNs with random weights. Then, we let the agent observe the state $s$ and take an action according to the initial main actor policy $\pi_\phi(s)$. The action is perturbed by adding Gaussian noise $\epsilon\sim\mathcal{N}(0,\sigma)$, with zero mean and variance $\sigma$. The computed action $a=\pi_\phi(s)+\epsilon$ is then applied to the transmitted waveforms, which are classified by the receiver that sends feedback to the transmitter. The Chares DRL agent extracts the new state $s'$ from the feedback and computes the reward $r$ (see Section \ref{sec:reward} for details on Chares reward mechanism). The tuple $(s,a,r,s')$, also known as \textit{trajectory}, is added to the experience buffer.

\begin{figure}[t]
    \centering
    \includegraphics[width=0.9\columnwidth]{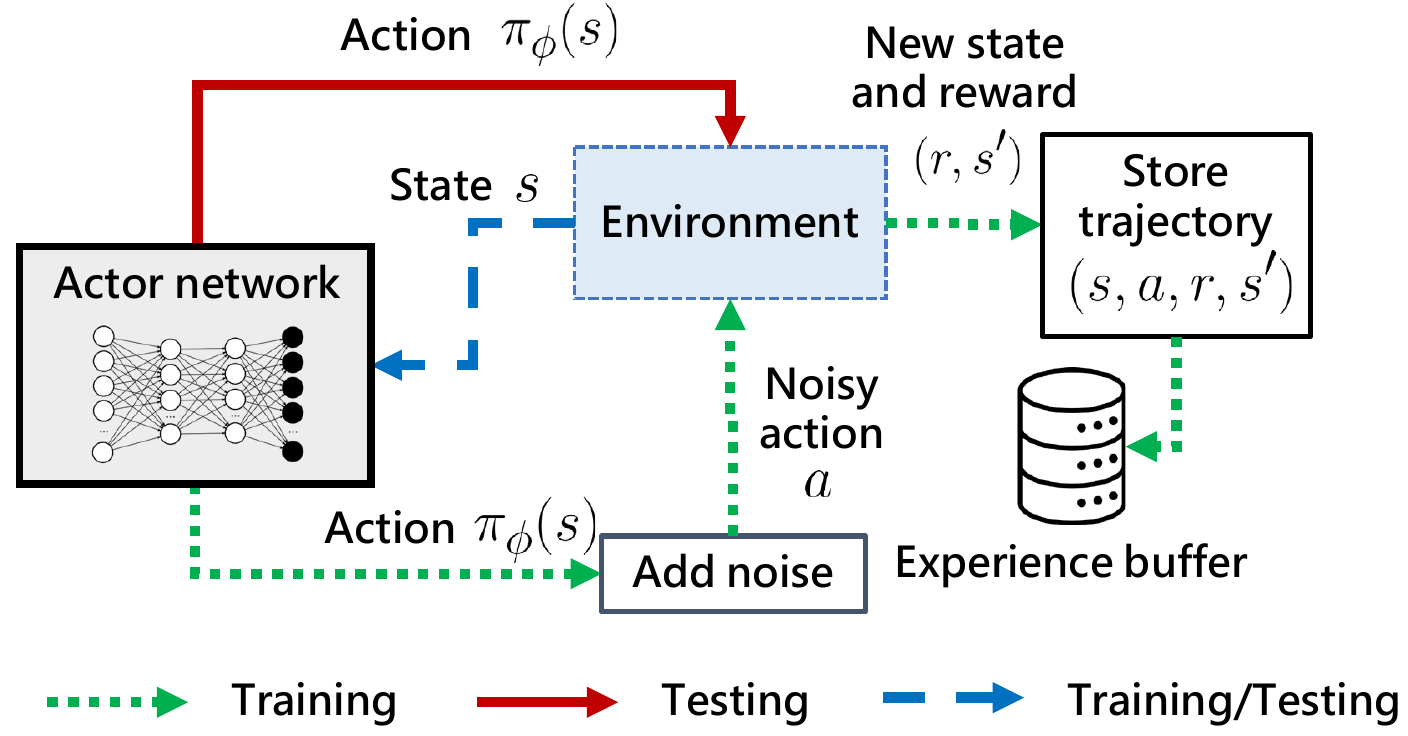}
    \vspace{-0.2cm}
    \caption{Actor network training and testing procedures.}
    \label{fig:actor}
\end{figure}

As shown in \reffig{fig:actor}, noise is added to actor policies during training only. At testing time, no noise is added to the policy. As soon as there are at least $B>0$ entries in the experience buffer, Chares DRL agent randomly extracts a batch of $B$ entries and, for each trajectory $(s_j,a_j,r_j,s'_j)$ in the batch, computes the noisy target actions $\tilde{a}_j = \pi_{\phi'}(s_j) + \epsilon$ and the target values
\begin{equation} \label{eq:target}
    y_j = r_j + \gamma \min_{i=1,2} Q_{\theta'_i}(s'_j,\tilde{a}_j)
\end{equation}

Target values in \eqref{eq:target} are used to update the weights $\theta_i$ of the main critic networks via stochastic gradient descent minimizing the mean-squared Bellman error (MSBE) loss function:

\begin{equation} \label{eq:loss}
    L_{\theta_i} = \frac{1}{B} \sum_{j=1}^B (Q_{\theta_i}(s_j,a_j) - y_j)^2
\end{equation}
\noindent
where the target values $y_j$ are computed as in \eqref{eq:target}.

MSBE is a common loss function used in DDPG architectures to measure how good approximated values generated by the critics are in satisfying the Bellman equation in \eqref{eq:q_val_nondet}, and the experience buffer helps critic networks in generating more reliable and accurate Q-value estimates. As shown in \reffig{fig:architecture}, target networks and the main actor network are updated every $d$ time-steps. Target networks are updated as follows:
\begin{align}
    \theta'_i & = \omega \theta_i + (1-\omega) \theta'_i \\
    \phi' & = \omega \phi_i + (1-\omega) \phi'_i
\end{align}
\noindent
which avoids abrupt updates of target weights (and thus stabilizes training procedures) by tuning the $\omega$ parameter taking values in $[0,1]$. Small $\omega$ values result in slow updates of the target weights, while $\omega$ values close to 1 rapidly copy main network weights onto target network ones. Finally, the main actor network weights $\phi$ are updated every $d$ steps via deterministic policy gradient \cite{ddpg} through gradient ascent \vspace{-0.3cm}

\begin{equation} \label{eq:det_pol_grad}
\phi^{t+1}\!=\! \phi^{t} \!+\!  \frac{\chi}{N} \sum_{j=1}^B \left [ \nabla_\phi \pi_\phi(s_j) \nabla_{a_j} Q_{\theta_1}(s_j,a_j)| a_j \!=\! \pi_\phi(s_j) \right], \nonumber
\end{equation}
\noindent
where $\chi$ is a (decreasing) step-size that ensures the convergence of the updates. Notice that while the main twin critics are updated by clipping Q-values from both networks, the main actor network is updated by using $\theta_1$ weights only. This step still guarantees convergence under mild assumptions \cite{td3}. \vspace{-0.3cm}

\subsection{Reward design} \label{sec:reward}

The reward system is aimed at incentivizing FIR taps $\mathbf{h}$ that increase the accuracy while penalizing those who result in worsened prediction results or decoding errors. Thus, a large reward $\rho^{SUCCESS}>0$ is given every time the feedback reports the correct classification label. A moderate reward $\rho^{UP}>0$ is given when the feedback shows better prediction results (\textit{e.g.}, the softmax output has increased with respect to the previous transmission). A negative reward $\rho^{DOWN}<0$ is given when the feedback shows that newly computed FIR taps have caused a degradation in either the accuracy of the classifier (\textit{e.g.}, wrong label or decreased softmax output) or the decoding success (\textit{e.g.}, high decoding failure rate). Finally, a reward $\rho^{SAME} = 0$ is given when the feedback shows that the system is performing the same as before\footnote{While this latter case is unlikely to happen in any real-world system, we have included it in our model to consider those cases where feedback is quantized to a finite set of discrete values.}. In Section \ref{sec:numerical}, we will demonstrate how this simple yet effective reward mechanism allows to achieve up to 4.1x gain when compared to other approaches.

 \vspace{-0.3cm}
\section{Experimental Results} \label{sec:numerical}

To demonstrate the effectiveness of Chares under diverse conditions, in this section we present results obtained by applying the Chares DRL agent to a variety of WSC problems and comparing its performance with existing approaches.

Our key objective is to show how Chares can be seamlessly ported from one WSC problem to another with minimum modifications. For this reason, we consider a unified architecture where critic and actor networks are implemented as fully-connected neural networks with 10 layers each consisting of 30 neurons with ReLU activation functions. The learning rate is set to $0.99$, target networks and main actor networks are updated every $d=2$ time-steps and weights are copied from main to target networks with parameter $\omega=0.05$. Unless otherwise stated, the number of taps is set to $M=11$ and the maximum deviation from the ideal FIR filter $\boldsymbol{\phi}^0=[1,0,0,\dots,0]$ is set to $\alpha=0.1$. The batch size used to extract trajectories from the experience buffer is set to $B=64$, while the buffer stores at most $10000$ entries. We consider the case where the receiver feeds back the classification label as well as the softmax output to the transmitter. The reward mechanism is setup as follows. Successful classification gives a reward $\rho^{SUCCESS}=2$, wrong classification but improved softmax output results in a reward $\rho^{UP}=1$, decreased softmax output gives a reward equal to $\rho^{DOWN}=-1$, and a reward $\rho^{SAME} = 0$ otherwise.

To showcase Chares capabilities, we consider two relevant WSC problems, \textit{i.e.}, modulation classification and RF fingerprinting, under three different configurations:

$\bullet$ \textit{Single-label (SLA):} this problem is relevant to RF fingerprinting WSC applications where a receiver must recognize a specific transmitter just by looking at small hardware impairments in the received waveforms, \textit{i.e.}, the features \cite{zheng2019fid,16-Peng-ieeeiotj2018,17-Xie-ieeeiotj2018,18-Xing-ieeecomlet2018}. In this case, Chares DRL agent is required to synthesize waveforms for the class identifying the transmitter;

$\bullet$ \textit{Multiple-labels (MLA):} this is the case where a transmitter changes modulation scheme over time and the receiver leverages DL to detect the employed modulation scheme and demodulate waveforms \cite{Kulin-ieeeaccess2018,deepsig,Karra-ieeedyspan2017,o2017introduction,xiong2019robust}. In this case, Chares must be able to compute FIR taps that are effective for diverse channel conditions and modulation schemes;

$\bullet$ \textit{Adversarial (ADV):} this configuration considers the case of an adversary injecting noise (e.g., a jammer) with the overall objective to produce misclassifications at the receiver side. Chares DRL agent must be able to counteract adversarial actions and ensure proper classification at the receiver side.

\vspace{-0.3cm}
\subsection{Datasets and DL models descriptions} \label{sec:numerical:datasets}

To train and test our DRL agent, we consider two wireless datasets for modulation classification and RF fingerprinting WSC problems. For modulation classification we use the publicly available DeepSig RADIOML 2018.01A dataset \cite{deepsig} containing waveforms from 24 different modulation schemes. For each modulation, the dataset provides approximately 106.000 waveforms under different SNR conditions from -20dB to +30dB. The classifier in this case is implemented via the CNN described in \cite[Table III]{deepsig}. The input of the classifier consists of a sequence of 1024 complex-valued IQ samples, and the classification is performed via majority-rule voting across a batch consisting of 32 waveforms. In \cite{deepsig}, authors have shown that the classifier achieves poor classification performance when the classifier is tested over low SNR conditions. For this reason, the classifier is trained with waveforms with high SNR (from 20dB to 30dB) values and then we let Chares operate under lower SNR conditions (-10dB to 20dB), thus simulating the case where the classifier is trained under controlled channel conditions, yet operates under noisy and fading channels.

The second dataset is a publicly available dataset tailored for RF fingerprinting applications \cite{shawabka2020exposing} containing waveforms recorded with a testbed of 10 Ettus Research USRP software-defined radios transmitting identical WiFi frames. Since in RF fingerprinting problems the DL model must be able to identify the transmitter from its hardware impairments only, frames are generated in GNUradio and contain the same MAC address, thus masking the real identity of the transmitter. In this case, we consider the Baseline CNN described in \cite[Section~3.B]{shawabka2020exposing} with a block consisting of two convolutional layers followed by ReLU and a MaxPool layer replicated 5 times, then followed by three fully connected layers. We focus on the single-antenna setup with equalized IQ samples (Setup D in \cite{shawabka2020exposing}) where waveforms are recorded at 10 different times of the day for 2 consecutive days. The input size is equal to 288 complex-valued IQ samples. As we will discuss in Section \ref{sec:num:sla}, we train the classifier on a specific time of day 1, but we test it with waveforms recorded at different times. This setup has been shown \cite{shawabka2020exposing} to challenge the classifier by bringing its accuracy close to random guessing. This setup accurately simulates the case shown in \reffig{fig:intro} where the classifier  operates under unseen channel conditions.\vspace{-0.2cm}

\begin{table}[t]
\caption{\label{tab:train_acc} Classification accuracy for different setups and problems}
\centering
\begin{tabular}{ccccc}
\hline
\multicolumn{5}{c}{Multi-label Classification (Modulation recognition \cite{deepsig}) - No FIR}                           \\
\hline
         & High SNR                & Low-Mid SNR                & Low SNR &                \\
\hline
BPSK     & 1                         & 0.94                   & 0.41               &                  \\
\hline
16QAM    & 0.68                      & 0.35                   & 0.11               &                  \\
\hline
64QAM    & 0.65                      & 0.63                   & 0.23               &                  \\
\\
\hline
\multicolumn{5}{c}{Single-label Classification (RF fingerprinting \cite{shawabka2020exposing})  - No FIR}                              \\
\hline
         & File 1 – Day 1                & All days                & Day 1                & Day 2                 \\
\hline
Device 7 & 1                         & 0.22                   & 0.31               & 0.15             \\
\hline
\end{tabular}
\end{table}

\subsection{Multi-label modulation classification} \label{sec:numerical:mla}

We first consider a realistic use case scenario where a WiFi transmitter implements adaptive modulation and coding scheme (MCS) by changing modulation according to time-varying channel conditions. The transmitter adaptively selects between MCS indexes 0,3 and 5, corresponding to BPSK, 16QAM and 64QAM. The receiver implements the aforementioned CNN classifier (Section \ref{sec:numerical:datasets}) which, among others, allows the receiver to detect the modulation of incoming waveforms and infer this knowledge to demodulate received packets. We assume that the classifier is trained with waveforms received in the high SNR regime (\textit{i.e.}, [16,30]dB) but after deployment, the receiver operates in the low (\textit{i.e.}, [-10,4]dB) to mid (\textit{i.e.}, [6,14]dB) SNR regimes. Table \ref{tab:train_acc} shows the classification accuracy of the classifier when operating in different SNR conditions. Notice that the accuracy is greater when testing on channel conditions that are similar to the ones experienced during training (\textit{i.e.}, high SNR), but plummets when operating in lower SNR regimes (up to 6$\times$ smaller).

 \begin{figure}[t!]
    \centering
    \includegraphics[trim=0 20 0 0,clip,width=\columnwidth]{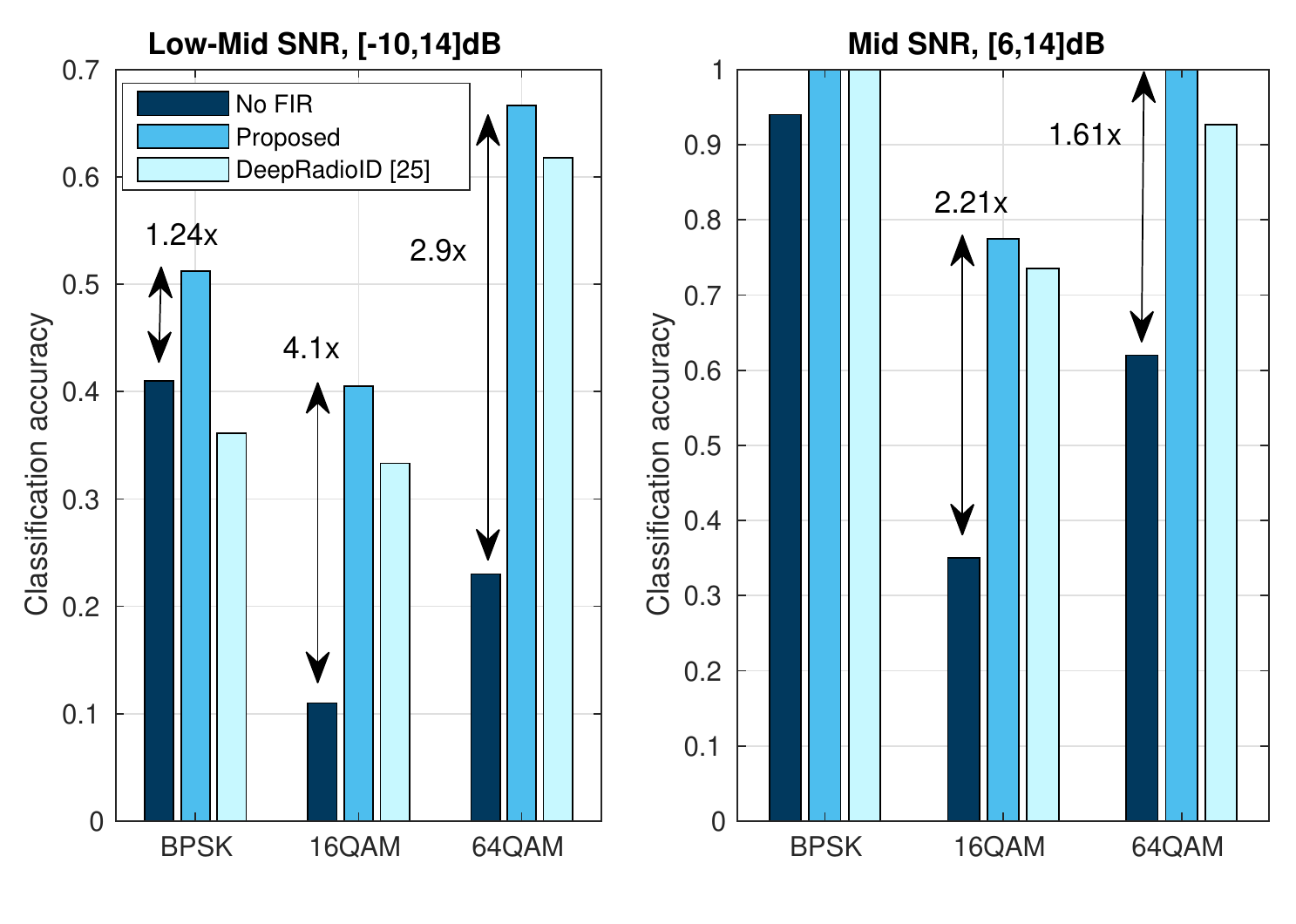}
    \caption{Classification accuracy of different approaches for the MLA problem.}
    \label{fig:acc_mod}
 \end{figure}

\reffig{fig:acc_mod} compares our solution against the baseline case with no waveform synthesis and DeepRadioID \cite{restuccia2019deepradioid}. We notice that Chares always outperforms both approaches -- when compared to the baseline, Chares provides accuracy improvements up to 4.1$\times$ with average improvements equal to 2.75$\times$ and 1.63$\times$ in the case of low-mid and mid SNR regimes, respectively. When compared with DeepRadioID, Chares improves the accuracy of the system by 1.1$\times$ on average. We point out that DeepRadioID (i) is trained offline for each class over the whole dataset, (ii) requires gradients of the classifier for each input, and (iii) computes one FIR filter to be used in all possible channel conditions. On the contrary, Chares is trained online and does not require \textit{any} information on the classifier and its architecture.

  \begin{figure}[t!]
    \centering
    \includegraphics[trim=45 20 45 20,clip,width=\columnwidth]{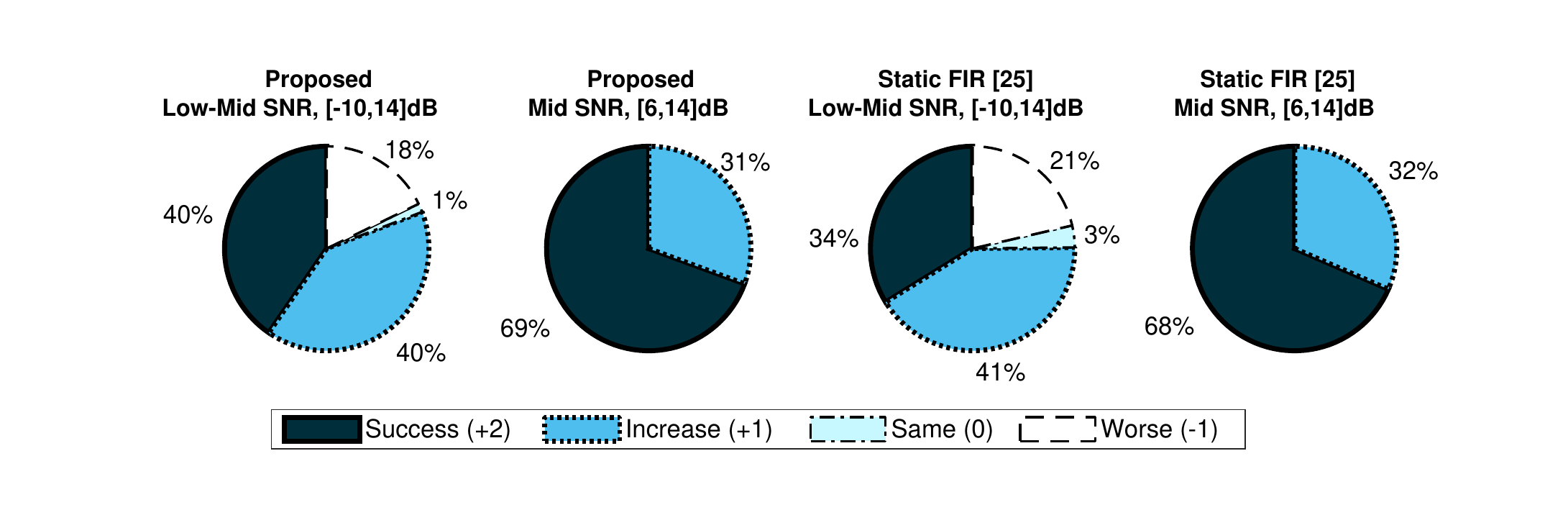}
    \caption{Reward distribution of different approaches for the MLA problem.}
    \label{fig:pie_mod}
 \end{figure}

\begin{figure}[b]
    \centering
    \includegraphics[width=0.9\columnwidth]{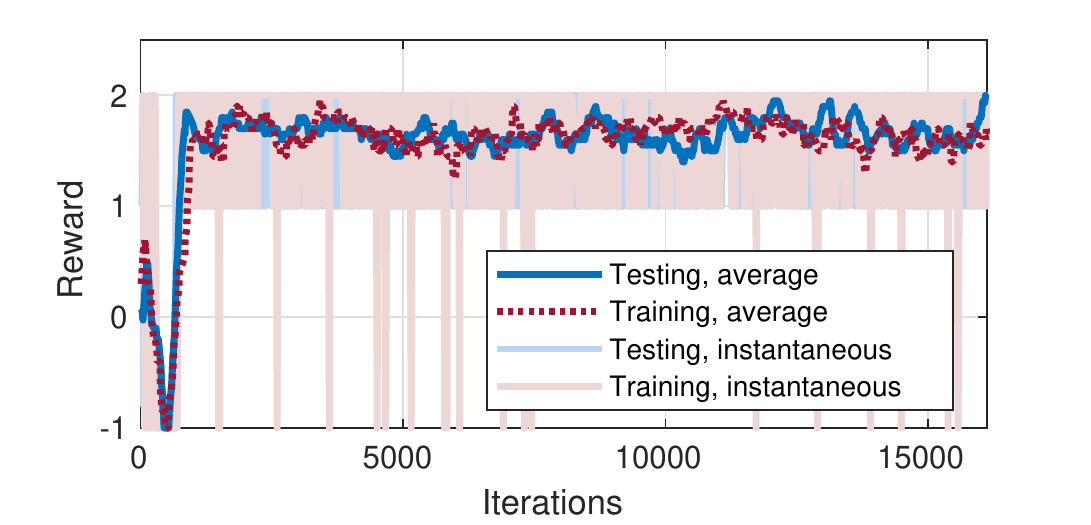}
     \vspace{-0.3cm}
    \caption{Reward of Chares for the MLA problem in the mid SNR case.}
    \label{fig:convergence}
 \end{figure}

To better understand how Chares impacts the classification process, \reffig{fig:pie_mod} shows how both Chares and DeepRadioID impact the output of the classifier. Although both solutions increase the softmax output of the classifier $40\%$ and $31\%$ of times under low-mid and mid SNR regimes respectively, Chares always provides a better success rate than DeepRadioID. Moreover, the latter generates FIR taps that result in higher softmax decrease rate in low-mid SNR regimes, while both solutions experience no decrease in performance when operating in mid SNR regimes. This demonstrates that using a unique FIR filter for different channel conditions is not an efficient solution, which shall be instead tackled with channel-specific approaches such as Chares.   Finally, \reffig{fig:convergence} shows the convergence speed of Chares DRL agent, specifically we show that Chares approaches the maximum reward $\rho^{SUCCESS}$ after approximately 1,000 learning iterations (see Section \ref{sec:arch:training} for a detailed breakdown of each training iteration).

 \vspace{-0.3cm}
\subsection{Single-label RF fingerprinting} \label{sec:num:sla}

Here, we have selected device 7 out of the ten devices in the dataset, as this device (see Table \ref{tab:train_acc}) has shown 100\% accuracy when trained and tested on day 1, but exhibits 15\% accuracy on average when tested with waveforms from day 2.

 \begin{figure}[t]
    \centering
    \includegraphics[width=0.95\columnwidth]{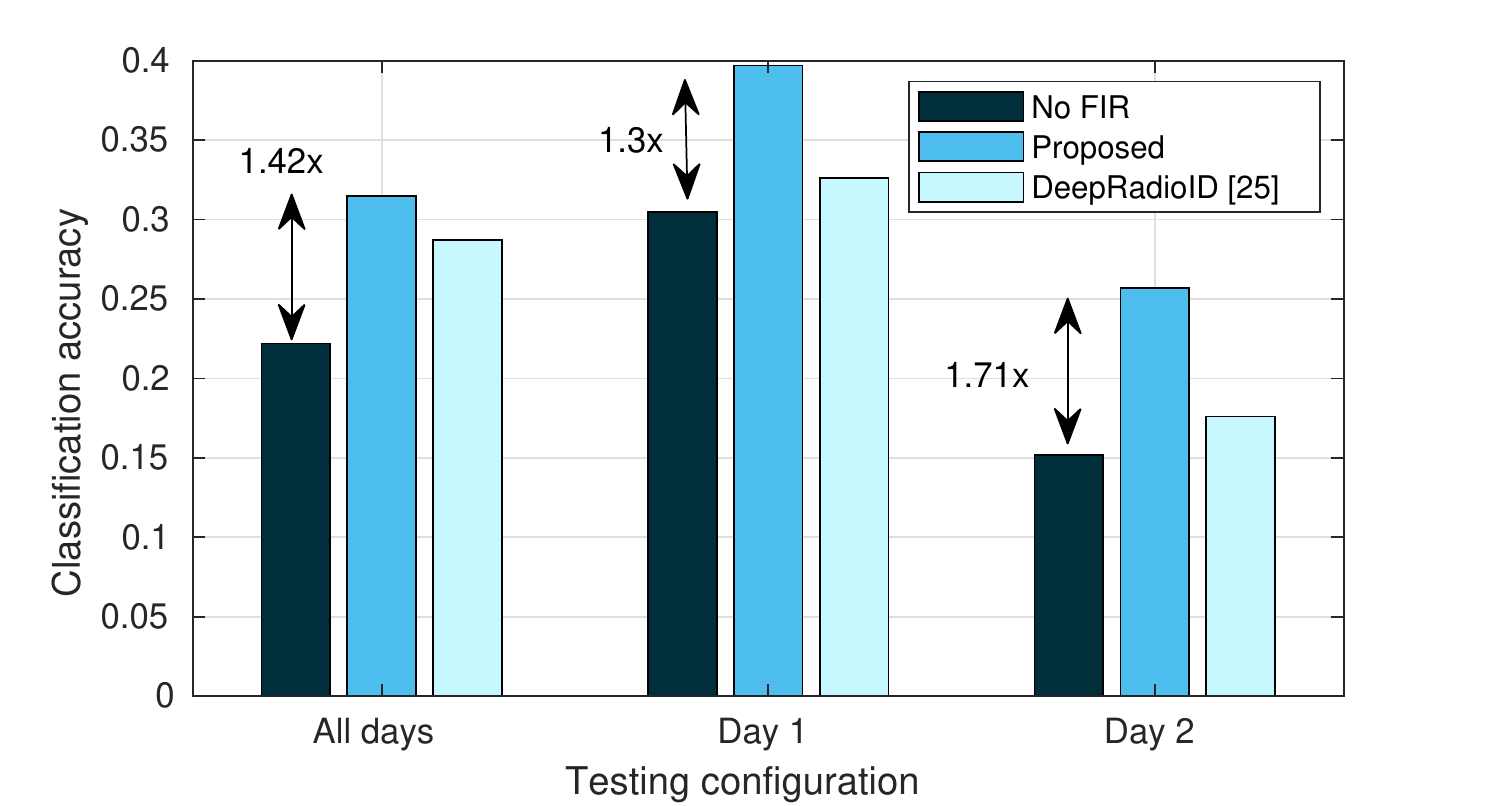}
    \vspace{-0.4cm}
    \caption{Comparison of classification accuracy of different approaches for the SLA problem.}
     \vspace{-0.3cm}
    \label{fig:acc_fing}
 \end{figure}

 \reffig{fig:acc_fing} shows the classification accuracy of device 7 for different setups. It is clear that the baseline classifier cannot generalize over different channel conditions. However, Chares was able to increase the accuracy up to factor 1.71$\times$ when tested on waveforms recorded on day 2. The reason is that although different, channel conditions during the same day are similar, meaning that the baseline classifier can still achieve a higher accuracy then the case where it is tested on a completely different day. As also illustrated in \reffig{fig:acc_fing}, Chares outperforms the state-of-the-art \cite{restuccia2019deepradioid} by effectively increasing the the success rate and providing an overall higher rewards.
   \begin{figure}[b!]
    \centering
    \includegraphics[width=0.9\columnwidth]{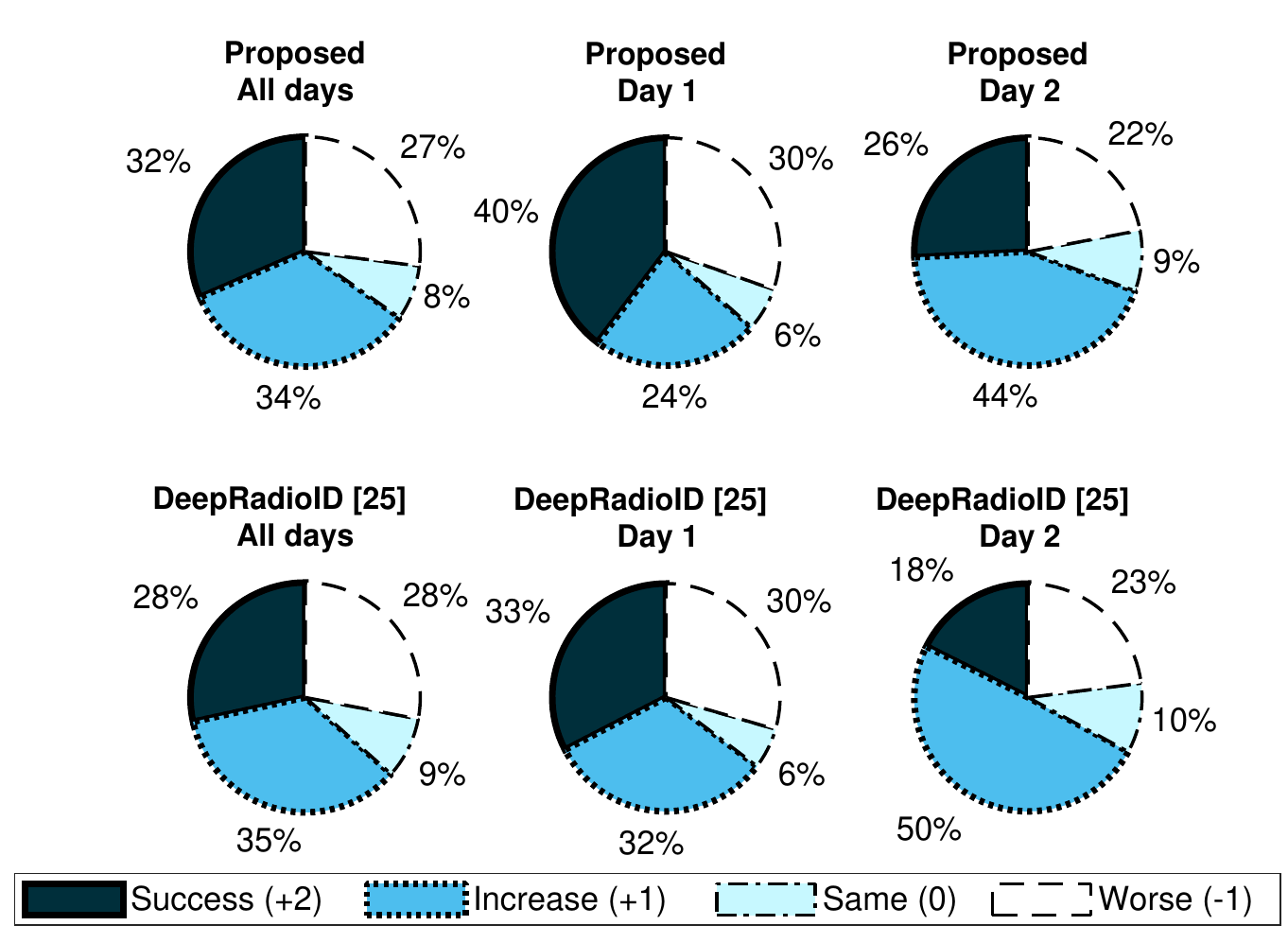}
    \caption{Comparison of reward distribution of different approaches for the SLA problem.}
    \label{fig:pie_fing}
 \end{figure}
 Notice that gains in the SLA case are lower than those achieved in the MLA case, as the RF fingerprinting dataset uses bitwise similar devices whose hardware impairments are extremely similar, which makes it hard for the classifier to distinguish between devices. In this case, the classifier is already prone to errors  due to the similarities between devices, and Chares can only improve the accuracy to a certain extent.\vspace{-0.4cm}

\subsection{Chares Adversarial Action Evaluation}

Finally, we analyze the case where a jammer transmits random waveforms that generate interference with those generated by the transmitter. In this case we have used the same Chares DRL agent and DeepRadioID models trained in the low-mid MLA case discussed in Section \ref{sec:numerical:mla} and we have tested them in the new adversarial environment. This use-case is probably the most relevant, as it shows how different waveform synthesis solutions perform over completely new and previously unseen channel conditions.
 \begin{figure}[t!]
    \centering
    \includegraphics[width=0.85\columnwidth]{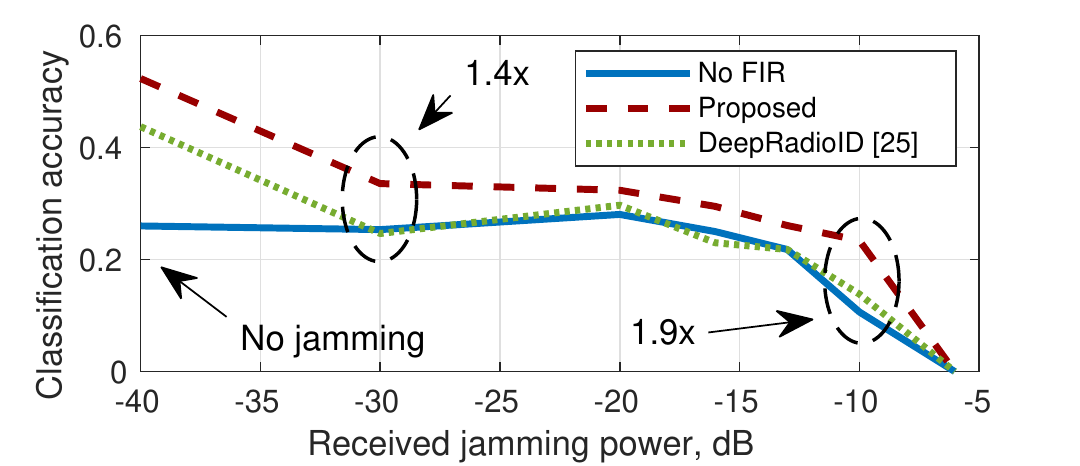}
    \vspace{-0.1cm}
    \caption{Classification accuracy in the case of an adversarial attack.}
    \label{fig:adversary}
 \end{figure}
\reffig{fig:adversary} shows the average classification accuracy of the three modulations for different solutions and received jamming power. In all cases, Chares outperforms both DeepRadioID and the case with no FIR filtering, by providing up to 1.9$\times$ accuracy increase when the jamming power is high, demonstrating how real-time and adaptive waveform synthesis effectively outperform offline and baseline approaches.

\subsection{Chares Real-Time Latency Analysis}

It is imperative to assess whether Chares can truly operate within typical coherence channel times. For this reason, we have synthesized the actor network of Chares (see Figure \ref{fig:actor}) in \gls{fpga}. We used high-level synthesis (HLS) to convert a C++-level description of the neural network in Verilog. Our target device is a  Xilinx xc7z045ffg900-2 \gls{fpga}, which is an \gls{fpga} used in typically used in software-defined radios such as USRPs. Our synthesis yields a latency of 13614 clock cycles with minimum clock period 3$\mu$s, which corresponds to 40.842 $\mu$s. As typical coherence channels are in the order of tens of milliseconds \cite{xie2018precise}, we are confident Chares can fully keep up with realistic channel dynamics. The resource utilization of our design is below 1\% of the total \gls{fpga} resources.

 \vspace{-0.3cm}
\section{Conclusions}\label{sec:conclusions}

We have presented Chares, an online DRL solution for channel-resilient resilient waveform synthesis applications. Chares leverages FIR filtering to synthesize transmitted waveforms, and can be trained in real-time independently from the specific WSC classifier model and objective. Moreover, Chares has been designed to adapt to unseen channel conditions and to be robust against noise and channel fluctuations. Our results have shown that Chares outperforms the state-of-the-art and increase the accuracy up to 4.1x if compared to baseline applications with no waveform synthesis and up to 1.9x if compared to other waveform synthesis solutions in the literature. We also demonstrate that Chares can execute with latency in the order of microseconds, which makes it ideal for real-time waveform synthesis.
\footnotesize
\bibliographystyle{IEEEtran}
\bibliography{francesco}

\end{document}